\documentclass[prb,floats,twocolumn,10pt]{revtex4}
\usepackage{graphicx,bm}

\begin{document}

\title {Magnons and Excitation Continuum in XXZ triangular antiferromagnetic model: Application to $Ba_3CoSb_2O_9$  }

\author{E. A. Ghioldi$^1$, A. Mezio$^1$, L. O. Manuel$^1$, R. R. P. Singh$^2$, J. Oitmaa$^3$, and A. E. Trumper$^1$}

\affiliation {$^1$Instituto de F\'{\i}sica Rosario (CONICET) and
Universidad Nacional de Rosario,
Boulevard 27 de Febrero 210 bis, (2000) Rosario, Argentina\\
$^2$Department of Physics, University of California, Davis, California 95616, USA\\
$^3$ School of Physics, The University of New South Wales, Sydney 2052, Australia}
\vspace{4in}
\date{\today}

\begin{abstract}
We investigate the excitation spectrum of the triangular-lattice antiferromagnetic $XXZ$ model using series expansion and mean field Schwinger boson approaches. The single-magnon spectrum computed with series expansion exhibits rotonic minima at the middle points of the edges of the Brillouin zone, for all values of the anisotropy  parameter in the range $0\leq J^z/J\leq1$. Based on the good agreement with series expansion for the single-magnon spectrum, we compute the full dynamical magnetic structure factor within the mean field Schwinger boson approach to investigate the relevance of the $XXZ$ model for the description of the unusual spectrum found  recently  in $Ba_3CoSb_2O_9$. In particular, we obtain an extended continuum above the spin wave excitations, which is further enhanced and brought closer to those observed in $Ba_3CoSb_2O_9$ with the addition of a second neighbor exchange interaction approximately  $15\%$ of the nearest-neighbor value. Our results support the idea that excitation continuum with substantial spectral-weight are generically present in two-dimensional frustrated spin systems and fractionalization in terms of {\it bosonic} spinons presents an efficient way to describe them. 
\end{abstract}
\maketitle

\section{Introduction}

The study of two dimensional (2D) quantum spin liquids (QSL) has been one of the central topics in  condensed matter physics.
Aided by frustration and strong quantum fluctuations,  novel quantum spin-liquid phases can emerge in quantum spin systems, which do not break any symmetry of the Hamiltonian and there is no local order parameter to unambiguously characterize them. Consequently, conventional paradigms of magnetism such as spin waves or Landau-Ginzburg-Wilson theories turn out to be inadequate.\cite{Anderson73,wen02,Senthil04,Misguich05,Sachdev08,balents10}
Several candidate materials showing such spin liquid behavior have been indirectly identified  by specific heat or nuclear relaxation measurements,\cite{powell11} however, a direct experimental detection of QSL remains elusive. \\
\indent Another signature of QSL is the emergence of spin-$\frac{1}{2}$ fractional excitations, called spinons. They have been predicted\cite{Fadeev81} in 1D AF's and detected\cite{Nagler91} by means of inelastic neutron scattering (INS) experiments.  Here, the spinon excitation is interpreted as a propagating domain wall; while the observed extended continuum in the spectrum is related to the different pairs of independently propagating spinons created in the AF system once spin-1 excitations are exchanged with the scattering of neutrons.\\
\indent In 2D the physical origin of spinons and their quantum statistics is more complex and not fully understood. However, it is widely believed that the extended continuum observed with INS in certain $2D$ compounds may also correspond to the fractionalization phenomenon.
Such a continuum has been observed in the inorganic compound\cite{Coldea01} $Cs_2CuCl_4$, the kagome-lattice Herbertsmithites $ZnCu(OH)Cl$,\cite{yslee} and
recently,\cite{Zhou12} in $Ba_3CoSb_2O_9$ which is an experimental realizations of the spin-$\frac{1}{2}$ triangular antiferromagnet, with very
little spatial distortion.\cite{Shirata12} In $Cs_2CuCl_4$ the superexchange interactions are spatially anisotropic\cite{Coldea01} while in $Ba_3CoSb_2O_9$ there is 
enough evidence of anisotropic spin spin interactions described by the $XXZ$ model in the easy-plane regime but little deviation from the triangular-lattice geometry.\cite{Susuki13,Batista13} 
While the Herbertsmithite materials remain disordered down to the lowest measured temperature, 
at sufficiently low temperatures, $Cs_2CuCl_4$ and  $Ba_3CoSb_2O_9$ are magnetically ordered, showing helical\cite{Coldea01} and $120^{\circ}$ long range N\'eel order,\cite{Zhou12} respectively. In the case of $Cs_2CuCl_4$ the spectrum shows  well defined magnon signals
at the Goldstone modes and a broad continuum with a dominant spectral weight at higher energies that persist even above the N\'eel temperature $T_N=0.62 K$. Due to the $2D$ character of the magnetic interactions this behavior was originally associated with the experimental realization of $2D$ spinons,\cite{Coldea01} however, further theoretical work showed that spinons in $Cs_2CuCl_4$ are actually of 1D character,\cite{Kohno07} that is, the combined effect of spatially anisotropic quantum fluctuations and frustration induce  an unexpected dimensional reduction.\cite{Zheng06II,Heidarian09,balents10} In contrast, though anisotropic in spin space, the magnetic interactions in $Ba_3CoSb_2O_9$ are $2D$ spatially isotropic. Therefore, the unusual broad and dominant continuum above the spin wave dispersion recently found below $T_N=3.8 K$ has been interpreted as a true 2D fractionalization, suggesting that the $120^{\circ}$ N\'eel phase  of this compound may be in close proximity to a spin liquid phase.\cite{Zhou12,Mourigal13}\\
 \indent In this article we investigate the energy spectrum of the triangular AF $XXZ$ model 
using two complementary techniques, series expansion (SE) and mean field Schwinger bosons (SBMF).\cite{Arovas88,Sarker89,Chandra90,Read91,Ceccatto93}  Series expansion gives reliable results for the dispersion relation of the single-magnon sector of the spectrum while a Schwinger bosons mean field allows us to study the whole energy spectrum with a spinon based theory through the dynamical magnetic structure factor.\cite{Arovas88}
In order to  take into account anisotropic exchange interactions within the SBMF approximation we have used four bond operators, as proposed by Burkov and Mac Donald within the context of 
quantum Hall bilayers.\cite{Burkov02} Our series expansion results show the presence of roton-like minima at the middle of the edges of the Brillouin zone that persist down to the $XY$ model. Motivated by the good agreement between series expansion and SBMF theory for the one magnon dispersion relation and based on the probable proximity of $Ba_3CoSb_2O_9$ to a spin liquid phase we 
study the effect of second neighbor exchange interactions on the whole spectrum of the $XXZ$ model using SBMF theory. In particular,  we find that a $15\%$  second neighbor interaction is enough to reproduce 
an extended continuum above the magnon excitations. For this frustration value there is a weakening of the attractive interaction between the two spinons building up the magnon excitation along with a strong reduction of the local magnetization. Therefore, our study 
provides a consistent calculation supporting the  recently proposed idea of fractionalization of $2D$ magnon excitations in $Ba_3CoSb_2O_9$.\cite{Zhou12}\\

The antiferromagnetic $XXZ$ model is defined as

\begin{equation} 
H= \sum_{\langle ij \rangle} [J (S^x_i S^x_j+S^y_i S^y_j)+ J^z S^z_i S^z_j],
\label{xxz}
\end{equation}
 
\noindent where the sum is over the nearest neighbor sites $\langle ij\rangle$ of a triangular lattice. This Hamiltonian breaks the $SU(2)$ symmetry of the Heisenberg model down to a $U(1)\times Z_2$.  In the easy plane case $0\leq J^z/J \leq 1$ and in the thermodynamic limit the $U(1)$ symmetry is broken by the ground state, selecting a $120^{\circ}$ N\'eel order state lying in the $x-y$ plane. This seems to be the case for the compound $Ba_3CoSb_2O_9$ below $T_N=3.8 K$. 

\section{Series expansion calculation}

To develop the series expansion, we first rewrite the Hamiltonian in a rotated basis, where the z axis points along the local spin direction of the $120^{\circ}$ ordered phase. \cite{Zheng06,book,review}
Then the Hamiltonian is written as $H_0 + \lambda V$, where $H_0$ consists of only Ising terms, with simple eigenstates and a ground state that
corresponds to one of the classical ground states. All other terms of the Hamiltonian are placed in $V$. The parameter
$\lambda$ is introduced artificially as an expansion parameter. The Hamiltonian of interest is realized at $\lambda=1$.  An additional ordering field term with coefficient $t(1-\lambda)$, with arbitrary $t$ is
used to improve the convergence of the series.\cite{book} Series expansion for the ground state energy and the order parameter for the $120^{\circ}$
N\'eel order were developed for various values of the anisotropy $\alpha=J^z/J$ complete to order $\lambda^{11}$. To analyze the magnetization
series, we perform a change of variables to remove a square-root singularity at $\lambda=1$\cite{book,review} and then develop d-log Pade approximants.
The estimated magnetization are shown in table I where the reduction of the zero point quantum fluctuations is clearly observed as anisotropy is increased.

The one-particle effective Hamiltonian is calculated as a power-series  in $\lambda$ for various real-space distances
from which the spectra at any wavevector
are readily obtained by Fourier transformation. These spectra are extrapolated to $\lambda=1$ by Pad\'e approximants. Calculations were done to order $\lambda^8$. We checked that the series for the Heisenberg
model ($J^z=J$) agreed completely with those obtained before.\cite{Zheng06} The SE results are shown in Fig. \ref{fig1} (magnenta dots) where   
%As always, the error-bars in the reported series expansion results are confidence intervals giving a measure
%of the spread in the different Pad$\'$e approximant values.
the predicted single-magnon spectrum shows the expected Goldstone mode structure at ${\bf k}\!=\!(0,0),\pm(\frac{4\pi}{3},0)$ (points O, Q and C of Fig. \ref{fig1}) for the isotropic case ($J^z=J$) and at ${\bf k}\!=\!(0,0)$ for the anisotropic case ($J^z < J $). Furthermore, the SE results exhibit roton-like minima at the middle points of the edges of the Brillouin zone, (point B of Fig. \ref{fig1}). Though not shown in the figure we found that the rotonic excitation persists down to the $XY$ model case ($J^z/J=0$).\cite{Chernyshev} These excitation should not be identitified to the local minima at momentum ${\bf k}\!=\!\pm(\frac{4\pi}{3},0)$ whose appearance
is  due to anisotropy effects. Originally, for the isotropic Heisenberg case, the rotonic excitations were described in terms of pairs of spinons\cite{Zheng06II} or vortex-antivortex\cite{Alicea06} excitations with fermionic character, or with conventional multi-magnon excitations in non-colinear antiferromagnets.\cite{Chernyshev,starykh-chubukov} Alternatively, it was  shown that the high entropy values found with high temperature expansion\cite{Eltsner93} could be reconciled by assuming a bosonic character for the rotonic excitations\cite{Zheng06} which within the Schwinger boson language can be interpreted as a pair of weakly bound spinon excitations\cite{Mezio12} (see below).

\section{Schwinger bosons for the XXZ model}
In this section, we further extend the widely used Schwinger boson representation to the $XXZ$ model. In contrast to the isotropic case\cite{Arovas88,Ceccatto93} and previous extensions\cite{Leone94,Fukumoto96} of the anisotropic case, we express the spin spin interaction in terms of four bond operators, as proposed by Burkov and Mac Donald,\cite{Burkov02} in order to preserve the original $U(1)\times Z_2$ symmetry of the Hamiltonian. Then, the  magnetically $120^{\circ}$ N\'eel order state that breaks the $U(1)$ symmetry is manifested by a Schwinger boson condensation which naturally occurs in the theory without assuming it from the beginning.\cite{Sarker89,Chandra90} 

Within the Schwinger bosons representation 
the spin operators components are written in terms of spin-$\frac{1}{2}$ bosons, $b_{\uparrow}$ and $b_{\downarrow}$ as

\begin{equation}
 \hat{S}^{x}_i=\frac{1}{2}(\hat{b}^{\dagger}_{i\uparrow}\hat{b}_{i\downarrow}+\hat{b}^{\dagger}_{i\downarrow}\hat{b}_{i\uparrow}),\;\;
 \hat{S}^{y}_i=\frac{1}{2\imath}(\hat{b}^{\dagger}_{i\uparrow}\hat{b}_{i\downarrow}-\hat{b}^{\dagger}_{i\downarrow}\hat{b}_{i\uparrow})
\end{equation}

\begin{equation}
 \hat{S}^{z}_i=\frac{1}{2}(\hat{b}^{\dagger}_{i\uparrow}\hat{b}_{i\uparrow}-\hat{b}^{\dagger}_{i\downarrow}\hat{b}_{i\downarrow}),\;\; \nonumber
\end{equation}

\noindent where the local constraint  

\begin{equation}
 \sum_{\sigma}
\hat{b}^{\dagger}_{i\sigma}\hat{b}_{i\sigma}=2s
\label{const}
\end{equation}

\noindent must be imposed to fulfill the spin algebra. The relevant bond operators for the $XXZ$ Hamiltonian Eq. (\ref{xxz}) are,

\begin{equation}
 \hat{A}_{ij}= \frac{1}{2} \left( b_{i\uparrow} b_{i\downarrow}-b_{i\downarrow}b_{j\uparrow} \right),\;\;\hat{B}_{ij}= \frac{1}{2} \left( b_{i\uparrow}b^{\dagger}_{j\uparrow}\!+\!b_{i\downarrow} b^{\dagger}_{j\downarrow} \right) \\ 
 \end{equation}
\begin{equation}
 \hat{C}_{ij}= \frac{1}{2} \left( b_{i\uparrow}  b^{\dagger}_{j\uparrow}-b_{i\downarrow} b^{\dagger}_{j\downarrow} \right),\;\;\hat{D}_{ij}= \frac{1}{2} \left( b_{i\uparrow} b_{j\downarrow} + b_{i\downarrow}b_{j\uparrow} \right).\\ 
\end{equation}

\noindent 
%Using the fact that under time reversal the spinor transforms as $(b_{i\uparrow},b_{i\downarrow})\rightarrow (b_{i\downarrow},-b_{i\uparrow})$ 
where $\hat{A}_{ij}$ and $\hat{B}_{ij}$ are $SU(2)$ and time reversal invariant while 
$\hat{C}_{ij}$ and $\hat{D}_{ij}$ are $U(1)$ invariant (rotation around $z$ axis) and change sign, $\small{\hat{C}_{ij}}\!\!\rightarrow\!\! -\hat{C}_{ij},\hat{D}_{ij}\!\!\rightarrow\!\! -\hat{D}_{ij} $, under time reversal.\cite{nota} Then, after writing down the spin operators in terms of Schwinger bosons, Eq.(\ref{xxz}) results\\

\begin{eqnarray} 
H=\frac{1}{2} \sum_{\langle ij\rangle} & [ (J + J^z) (:\!\hat{B}_{ij}^\dag \hat{B}_{ij}\!:-\hat{A}_{ij}^\dag \hat{A}_{ij}\!)- \nonumber \\
&-(J-J^z)(:\!\hat{C}_{ij}^\dag \hat{C}_{ij}\!: - \!\hat{D}_{ij}^\dag \hat{D}_{ij}\!)].
\label{xxzsb}
\end{eqnarray}

\noindent Noticing that the inversion of $S_{i}^z$ can be carried on  as a time reversal operation followed by a $\pi$ angle rotation around $z$ axis, it is easy to check that the original $U(1) \times Z_2$ symmetry of the $XXZ$ model is preserved by Eq.(\ref{xxzsb}).

\subsection{Mean field approximatiom}
Now a non trivial mean field decoupling of Eqs. (\ref{xxzsb}) can be implemented as, 

\begin{equation}
 \small{\hat{X}_{ij}^\dag \hat{X}_{ij}\approx \langle\hat{X}_{ij}^\dag\rangle \hat{X}_{ij} +\hat{X}_{ij}^\dag  \langle \hat{X}_{ij} \rangle - \langle \hat{X}_{ij}^\dag \rangle \langle \hat{X}_{ij} \rangle },
\label{mf}
\end{equation}

\noindent where $\hat{X}=\hat{A},\hat{B},\hat{C},$ and $\hat{D}$. From all the possible Ansatze  we choose translational invariant mean field parameters such as  $\langle \hat{A}_{ij} \rangle=\imath A_{ij}, \langle \hat{C}_{ij} \rangle =\imath C_{ij}$, $\langle \hat{B}_{ij} \rangle = B_{ij}$, and $ \langle \hat{D}_{ij}\rangle=D_{ij}$  with $A_{ij}= -A_{ji}$, $C_{ij}=-C_{ji}$, $B_{ij}=B_{ji}$, and $D_{ij}=D_{ji}$ all real. In principle, the resulting mean field Hamiltonian $H_{MF}$ breaks the time reversal symmetry which followed by the 
$\pi$ angle rotation around $z$ realizes the $S^z_i$ inversion. So, the $Z_2$ symmetry seems to be broken. However, if $H_{MF}$ is gauge transformed  as $G_T^{-1}H_{MF}G_T=H^{\prime}_{MF}$, where $G_T: b_{\sigma}\rightarrow  b_{\sigma} e^{-\imath \frac{\pi}{4}} $, time reversal symmetry is restored by $H^{\prime}_{MF}$  and consequently the $Z_2$ symmetry is also preserved. 
Actually we choose the above Ansatze because in the thermodynamic limit it is compatible with the semiclassical $120^{\circ}$ N\'eel state lying in the $x-y$ plane.\cite{ansatz_mf} 
Replacing Eq. (\ref{mf}) in Eq. (\ref{xxzsb}) and  following the standard procedure\cite{Mezio11} we arrive to the diagonalized mean field Hamiltonian

 \begin{equation}
 \small{\hat {H}_{\scriptscriptstyle MF} =  \sum_{\bf k} \omega_{{\bf k} \uparrow} \alpha^{\dagger}_{{\bf k} \uparrow} \alpha_{{\bf k} \uparrow} + \omega_{-{\bf k} \downarrow} \alpha^{\dagger}_{-{\bf k} \downarrow} \alpha_{-{\bf k} \downarrow} + E_{\scriptscriptstyle MF}} 
\end{equation}

\noindent with the spinon relation dispersion defined as

\begin{equation}
 \omega_{{\bf k} \uparrow}=\omega_{-{\bf k} \downarrow}=\omega_{\bm k} = \sqrt{[\Gamma^{BC}_{\bf k}+\lambda]^2 - [\Gamma^{AD}_{\bf k}]^2},
\label{disp}
\end{equation}

\noindent with

$$
\Gamma^{BC}_{\bf k}= \frac{1}{2}(1+\frac{J^z}{J})\gamma_{\bf k}^{\scriptscriptstyle B} - \frac{1}{2}(1-\frac{J^z}{J}) \gamma_{\bf k}^{\scriptscriptstyle C}
$$
$$
\Gamma^{AD}_{\bf k}= \frac{1}{2}(1+\frac{J^z}{J}) \gamma_{\bf k}^{\scriptscriptstyle A}- \frac{1}{2}(1-\frac{J^z}{J}) \gamma_{\bf k}^{\scriptscriptstyle D}
$$

\noindent and  
$$
\gamma_{\bf k}^{\scriptscriptstyle A} = \sum_{{\bm \delta} >0} \! J A_{\bm \delta} \sin ({\bf k}\! \cdot \!\bm{\delta}) \; \;\;\;\;\gamma_{\bf k}^{\scriptscriptstyle B} = \sum_{{\bm \delta} >0} \! J B_{\bm \delta} \cos({\bf k} \!\cdot \!{\bm \delta}) \\
$$

$$
\gamma_{\bf k}^{\scriptscriptstyle C} = \sum_{{\bm \delta} > 0} \! J C_{\bm \delta} \sin ({\bf k} \! \cdot \! {\bm \delta}) \; \; \; \;\; 
\gamma_{\bf k}^{\scriptscriptstyle D} = \sum_{{\bm \delta} > 0} \! J D_{\bm \delta}  \cos ({\bf k} \! \cdot \! {\bm \delta}),
$$

\noindent where $\bm{\delta}= {\bf r}_j-{\bf r}_i$ are the vectors connecting the first neighbors of the triangular lattice. The ground state mean field energy results

\begin{eqnarray}
E_{\scriptscriptstyle MF} &=& \frac{1}{2}\sum_{\bf k} \omega_{\bf k} - \lambda (2S+1) N=  \\ 
&=&3N\left[ (J+J^z) (B^2_{\bm \delta}-A^2_{\bm \delta})-(J-J^z)  (C^2_{\bm \delta}-D^2_{\bm \delta}) \right]. \nonumber
\end{eqnarray}

\noindent Notice that $\lambda$ is the Lagrange multiplier introduced to enforce, on average, the local constraint of Eq. (\ref{const}).

\indent The self consistent mean field equations are

\begin{eqnarray}
 S+{\textstyle \frac{1}{2}}\! \!& = &\!\! \frac{1}{2N}  \sum_{\bf k} \frac{\Gamma^{BC}_{\bf k}+\lambda}{\omega_{\bf k}}  \label{ec_aut_1}\\
 {A}_{\bm \delta} \!\!& = &\!\! \frac{1}{2N} \sum_{\bf k}  \frac{\Gamma^{AD}_{\bf k}}{\omega_{\bf k}} \sin \left( {\scriptstyle {\bf k} \cdot \bm{\delta}}\right) \label{ec_aut_2}\\
 {B}_{\bm \delta} \!\!& = &\!\! \frac{1}{2N} \sum_{\bf k}  \frac{\Gamma^{BC}_{\bf k}+\lambda}{\omega_{\bf k}} \cos \left( {\scriptstyle {\bf k} \cdot \bm \delta} \right)  \label{ec_aut_3} \\
  {C}_{\bm \delta} \!\!& = &\!\! \frac{1}{2N} \sum_{\bf k} \frac{\Gamma^{BC}_{\bf k}+\lambda}{\omega_{\bf k}} \sin \left( {\scriptstyle {\bf k} \cdot \bm \delta} \right) \label{ec_aut_4} \\
 {D}_{\bm \delta}\!\! & = &\!\! \frac{1}{2N} \sum_{\bf k}  \frac{\Gamma^{AD}_{\bf k}}{\omega_{\bf k}} \cos \left( {\scriptstyle {\bf k} \cdot \bm \delta} \right). \label{ec_aut_5}
\end{eqnarray}

\noindent As we have pointed out the present mean field approximation preserves the original $U(1)\times Z_2$ symmetry of the $XXZ$ Hamiltonian. Nonetheless, it turns out that the minimum of the spinon dispersion at $\frac{\bf Q}{2}$ behaves as $\omega_{\frac{\bf Q}{2}} \rightarrow 1/N$, implying the occurrence of a Bose condensation of $\hat{b}_{\uparrow}$ and $\hat{b}_{\downarrow}$ at ${\bf k}=\frac{\bf Q}{2}$ and ${\bf k}=-\frac{\bf Q}{2}$, respectively, in the thermodynamic limit (see Eq. (\ref{disp})). This is interpreted as the rupture of the continuous $U(1)$ symmetry. In particular, by working out the static structure factor the singular mode, ${\bf k}=\frac{\bf Q}{2}$, of Eqs.(\ref{ec_aut_1})-(\ref{ec_aut_5}) can be simply related to the local magnetization $m$ and, after converting the sums into integrals, it is obtained a new set of self consistent equations. corresponding to the thermodynamic limit.\cite{Sarker89,Chandra90,Mezio13} Alternatively, there is another way to compute the local magnetization $m$, that we have checked to be  completely equivalent to the previous one, which consists of solving the self consistent Eqs. (\ref{ec_aut_1})-(\ref{ec_aut_5}) for finite size systems and then perfom a size scaling  of the expression,\cite{Mezio11}

\begin{equation}
 m=\frac{1}{2N} \frac{\Gamma^{BC}_{\frac{\bf Q}{2}}+\lambda}{\omega_{\frac{\bf Q}{2}}},
\label{magfinite}
\end{equation}
        
\noindent which in the thermodynamic limit corresponds to the singular mode of Eq. (\ref{ec_aut_1}) when ${\bf Q}=(\frac{4\pi}{3},0)$ is the magnetic wave vector of the $120^{\circ}$ N\'eel order.  In table I is shown the local magnetization $m$ predicted by the SBMF for several anisotropy values resulting from the extrapolation of Eq. (\ref{magfinite}) in the thermodynamic limit. The predictions of series expansion and linear spin wave theory are also shown for comparison. It is worth to stress that 
the SBMF predictions compare quite well with that of series expansion as soon as anisotropy is increased. 

%Notice that in contrast to the isotropic case,\cite{Mezio13} where the condensation of both flavors occurs at ${\bf k}=\pm \frac{\bf Q}{2}$, in the $XXZ$ model the condensation only occurs at ${\bf k}= +\frac{\bf Q}{2} $, opening an energy gap at ${\bf k=-\frac{\bf Q}{2}}$.

\begin{table}[h]
\begin{tabular}[c]{ccccccc}
\hline
\hline
 $J^z/J$         &$\;\;$ &  SBMF     &$\;\;$ &  LSWT  &$\;\;$ & Series     \\
\hline
\hline
$1 $            && $0.2739$  && $0.2386$  && $0.198\pm 0.034$        \\
$ 0.8$          &&$0.3402$   && $0.3522$  && $0.245\pm 0.026$        \\
$0.6$           &&$0.3663$   && $0.3858$  && $0.283\pm 0.023$        \\
$0.4$           &&$0.3862$  &&  $0.4096 $ && $0.314\pm 0.018$       \\
$0.0$           &&$0.4204$   &&  $0.4485$ && $0.403\pm 0.005$          \\
\tableline
\end{tabular}
\caption{Local magnetization $m$
of the $120^{\circ}$ N\'eel ground state of the spin-$\frac{1}{2}$ antiferromagnetic $XXZ$  model on the triangular lattice obtained
within  the present mean field Schwinger bosons (SBMF), the linear spin wave theory (LSWT) and Series expansion.}
\label{table1}
\end{table}

\subsection{Dynamical structure factor}

In this subsection we study the spectrum by computing the $zz$ component of the spin spin dynamical structure factor,
 $\small{S^{zz}({\bf k},\omega)\!=\frac{1}{2 \pi} \int^{\infty}_{\infty} \langle S^z_{\bf k}(t) S^z_{-\bf k}(0)\rangle \exp^{\imath \omega t} dt }$.
The computation and the interpretation of the spectrum is based on our previous work performed for the isotropic case.\cite{Mezio11}     
Here we also work on finite systems so the continuous $U(1)$ symmetry is, in principle, preserved. However, one can access to the thermodynamic limit by extrapolating from finite size systems, as we previously did for the local magnetization study (Sec. III A). In fact, as soon as the long range $120^{\circ}$ N\'eel order is developed in the $x-y$ plane the spectrum corresponding to the transversal spin-$1$ excitations can be obtained. Even if the calculation is similar to the isotropic case we consider appropriate to outline again the main steps in order to develop a self contained subsection and to point out the differences that turn out for the $XXZ$ case. 
Following references\cite{Mila90,Lefmann94, Mezio11} the dynamical structure factor within the mean field Schwinger bosons results
\begin{equation}
S^{zz}({\bf k},\omega)\!=\!\frac{1}{4N}\sum_{{\bf q}} (u_{\bf q} v_{{\bf k}+{\bf q}} - u_{{\bf k}+{\bf q}} v_{\bf q})^2 
\delta (\omega-(\omega_{{\bf q}\uparrow}+\omega_{{\bf k}-{\bf q} \downarrow})),
\label{Skw}
\end{equation}

\noindent where \small{$u_{\bf k}\!\!= \!\![\frac{1}{2}(1\!+\!\frac{\Gamma^{BC}_{\bf k}+\lambda}{\omega_{\bf k}} )]^{\frac{1}{2}}$} and  \small{$v_{\bf k}\!=\!\!{\rm sgn}(\Gamma^{AD}_{\bf k})[\frac{1}{2}(-1+\frac{\Gamma^{BC}_{\bf k}+\lambda}
{\omega_{\bf k}} )]^{\frac{1}{2}}$} are the coefficients of the Bogoliubov transformation that diagonalizes $\hat{H}_{\rm MF}$. $S^{zz}({\bf k},\omega)$ consists of two free spinon excitations that give rise to a continuum. However, two distinct contributions can be identified in the spectrum, 

\begin{equation}
 S^{zz}({\bf k},\omega)=S_{sing}^{zz}({\bf k},\omega)+S_{cont}^{zz}({\bf k},\omega),
\end{equation}

\noindent where the singular part $S_{sing}^{zz}({\bf k}, \omega)$ represents the process of destroying one spinon $b_{-\frac{\bf Q}{2} \downarrow}$ ($b_{\frac{\bf Q}{2} \uparrow}$) of the condensate and creating another one $b^{\dagger}_{{\bf k}+\frac{\bf Q}{2} \uparrow}$ ($b^{\dagger}_{{\bf k}-\frac{\bf Q}{2} \downarrow}$) in the normal fluid, while the continuum part 
$S_{cont}^{zz}({\bf k},\omega)$ corresponds to the process of creating two spinons in the normal fluid only. 
Using the fact that 
$u_{\frac{\bf Q}{2}}= v_{\frac{\bf Q}{2}}\sim ({Nm})^{\frac{1}{2}}$ and $\omega_{\frac{\bf Q}{2} \uparrow}=\omega_{\frac{\bf Q}{2} \downarrow}\sim 0$,
the singular part can be approximated as 

\small{
\begin{eqnarray}
 S_{sing}^{zz}({\bf k},\omega)&\approx &\frac{m}{4} ( v_{{\bf k}+\frac{\bf Q}{2}}-u_{{\bf k}+\frac{\bf Q}{2}})^2 
\delta(\omega-\omega_{{\bf k}+\frac{\bf Q}{2}\uparrow}) + \nonumber \\
&&+\frac{m}{4} (u_{\!\frac{\bf Q}{2}-{\bf k}\!}\! -\! v_{\!\frac{\bf Q}{2}-{\bf k}\!})^2 \delta (\omega-\omega_{{\bf k}-\frac{\bf Q}{2} \downarrow}\!),
\label{singular}
\end{eqnarray}
}

\noindent while the continuum part results very similar to Eq. (\ref{Skw})
\begin{equation}
S_{cont}^{zz}({\bf k},\omega)\!=\!\frac{1}{4N}\sum_{{\bf q}}{}^{'} (u_{\bf q} v_{{\bf k}+{\bf q}} - u_{{\bf k}+{\bf q}} v_{\bf q})^2 
\delta (\omega-(\omega_{{\bf q}\uparrow}+\omega_{{\bf k}-{\bf q} \downarrow})),
\label{Skwp}
\end{equation}

\noindent except that in the sum over the triangular Brillouin zone  the momentum ${\bf q}$ satisfying ${\bf q}=\frac{\bf Q}{2}$ or 
${\bf k}+{\bf q}=\frac{\bf Q}{2}$ are not taken into account. This is indicated by the primed sum.\\
 
From Eq. (\ref{singular}) it is clear that the spectral weight of the singular part is located at the shifted spinon excitations $\omega_{{\bf k}+\frac{\bf Q}{2}\uparrow}$ and 
$\omega_{{\bf k}-\frac{\bf Q}{2} \downarrow}$. However, we have recently shown\cite{Mezio11} that --due to the coefficients in front of each delta function-- the spectral weight between both shifted spinon dispersions is redistributed in such a way that if one reconstructs a new dispersion from those pieces of the spinon dispersions with dominant spectral weight, the main features of the {\it one magnon} dispersion computed with the series expansion are recovered\cite{Zheng06II} (see, for instance in Fig. {\ref{fig1}} the $J^z=1$ case). Notice that for the anisotropic case $S_{sing}^{zz}({\bf k},\omega)$ does not contain elastic processes at 
${\bf k}=\pm{\bf Q}$. This is in contrast to the isotropic case where the condensation of up/down flavors occurs at both momenta, $\frac{\bf Q}{2}$ and $-\frac{\bf Q}{2}$. On the other hand, we have shown\cite{Mezio11} that the remaining weak signal dispersion is related to unphysical excitations coming from the relaxation of the local constraint. In fact, to recover the proper low temperature behavior of thermodynamic properties such unphysical excitations must be discarded.\cite{Mezio12} So, this simple procedure can be conceived as an approximate manner of carrying on the projection of the spectrum into the physical Hilbert space which, even numerically,\cite{Sorella12} is very difficult to implement in a calculation. We have called this procedure of reconstructing the one magnon excitation and eliminating the remnant weak signal {\it reconstructed} mean field Schwinger boson theory.\cite{Mezio12} 
\newpage
\begin{widetext}
\begin{center}
{\begin{figure}[t] 	 	
\vspace*{0.cm}
\includegraphics*[width=0.8\textwidth]{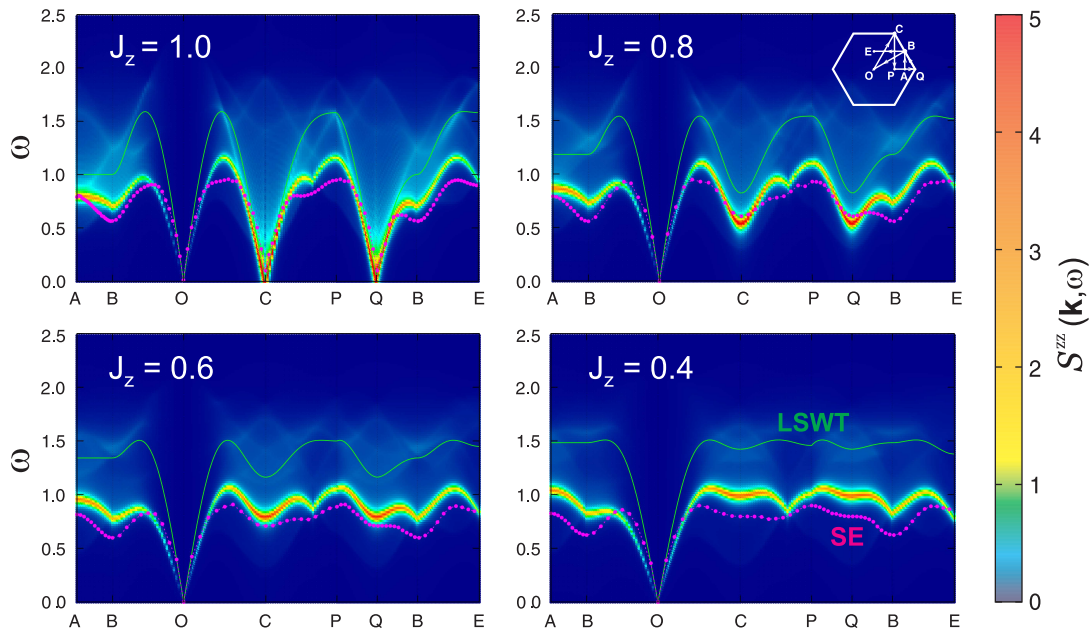}
\caption{Relation dispersion predicted by series expansion (magenta dots) and dynamical structure factor computed within the reconstructed SBMF theory (intensity curves),  along the path of the Brillouin zone shown in the inset, for several anisotropy values. The LSWT results (thin green line) are shown for comparison. Inset: path of the Brillouin zone, O$=(0,0)$, P=$(\frac{2\pi}{3},0)$, A$=(\pi,0)$, B$=(\pi,\frac{\pi}{\sqrt{3}})$, C$=(\frac{2\pi}{3},\frac{\pi}{\sqrt{3}})$, Q$=(\frac{4\pi}{3},0)$, and E$=(0,\frac{\pi}{\sqrt{3}})$. The values of $J_z$ are in units of $J$.}
\label{fig1}
\end{figure}}
\end{center} 
\end{widetext}
  
In Fig \ref{fig1} is shown the dynamical structure factor (intensity curve) within the reconstructed SBMF theory for different anisotropy values  $J^z/J$ along with the relation dispersion predicted by  series expansion (SE) and linear spin wave theory (LSWT). 
It is observed that the low energy sector of the spectra  predicted by the reconstructed SBMF theory reproduces quite well, qualitatively and quantitatively, the one magnon dispersion predicted by series expansion for the anisotropy range $0.4\lesssim J^z/J\leq1$. This agreement between the SBMF theory and SE  along with  
that obtained for the local magnetization (table I) give a strong support to the Schwinger boson mean field theory developed in section III for the $XXZ$ model.\cite{otros}  

\section{Application to $Ba_3CoSb_2O_9$ }

In this section we explore the possible relation between the present spectrum of the $XXZ$ model and that found in the INS experiments of $Ba_3CoSb_2O_9$. One important difference is that within the reconstructed SBMF the dominant spectral weight is mostly located at the low energy sector of the spectrum. However, given the proximity to a spin liquid phase proposed in the literarture,\cite{Zhou12} it is important to investigate the spectrum once the ground state of the $XXZ$ model is pushed  near a spin liquid phase. In our approximation this situation can be induced by introducing exchange interactions to second neighbors. In fact, in the isotropic case, it is known\cite{Kaneko14,Manuel99} that there is a spin liquid phase for moderate $J_2$ values, $0.1\leq J_2/J\leq 0.14$. Around these  $J_2$ values, and for small anisotropy $J^z_2/J_2=J^z/J=0.8$,  we have checked that the local magnetization is still quite robust but it is proximate to a spin liquid phase since it vanishes abruptly at $J_2/J \sim 0.25$.   
In Fig. \ref{fig2} is shown the dynamical structure factor (intensity curve) for several values of $J_2$. As $J_2$ increases there is an important spectral weight transfer from the low to the high energy sector of the spectrum. In particular, around $J_2/J= 0.15$  the extended continuum of two spinon excitations is recovered.\\ 
\indent As at the mean field level the spectrum corresponds to two free spinon excitations, it is important to get some insight about the spinon spinon interaction once corrections to
the mean field theory are included. Effective gauge field theory\cite{Read91,Sachdev08} predicts that for a commensurate spinon condensed phase there is a confinement of spinons, giving rise to spin-$1$ magnon excitations of the $120^{\circ}$ N\'eel order. Within the context of the Schwinger bosons one should include Gaussian fluctuations\cite{Trumper97} of the mean field parameters which is beyond 
the scope of the present work. Instead,  we adopt a simpler strategy\cite{Mezio12} that allows us to get a physical insight about the spinon spinon interactions once $J_2$ is included.

\begin{widetext}
\begin{center}
{\begin{figure}[t]
\vspace*{0.cm}
\includegraphics*[width=0.8\textwidth]{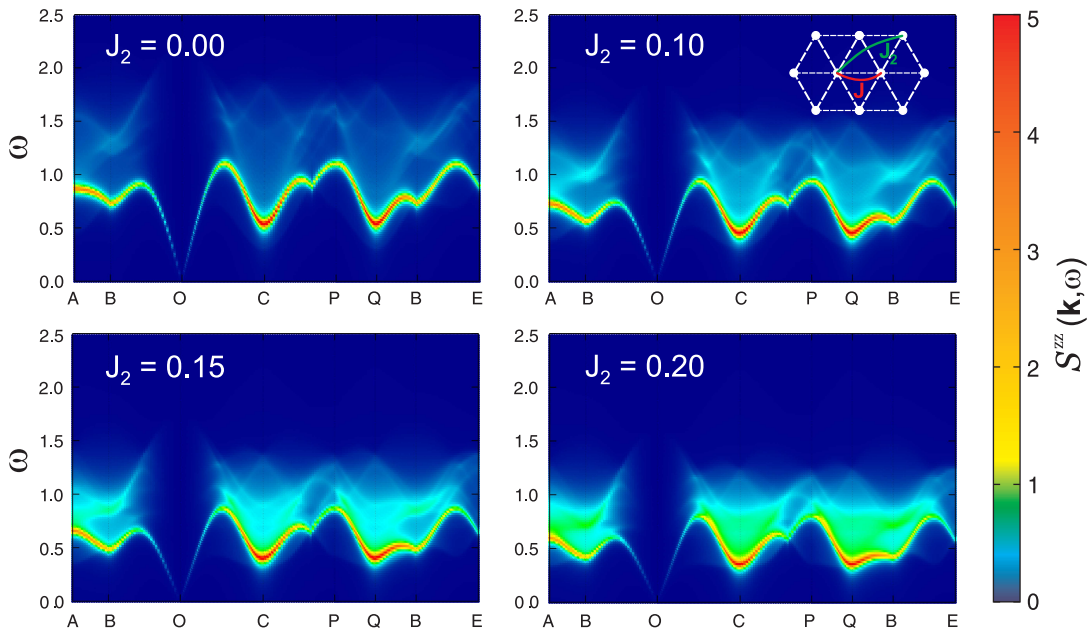}
\caption{Dynamical structure factor predicted by the reconstructed SBMF theory (Intensity curves) along the same path of the Brillouin zone of Fig. \ref{fig1} for several second neighbors exchange values. The same anisotropy interaction $J^z/J=0.8$, between first and second neighbors, has been selected. Inset: Schematic representation of the exchange interactions on the triangular lattice. The values of $J_2$ are in units of $J$.}
\label{fig2}
\end{figure}}
\end{center} 
\end{widetext}

\begin{figure}[h]
\vspace*{0.cm}
\includegraphics*[width=0.40\textwidth,angle=0]{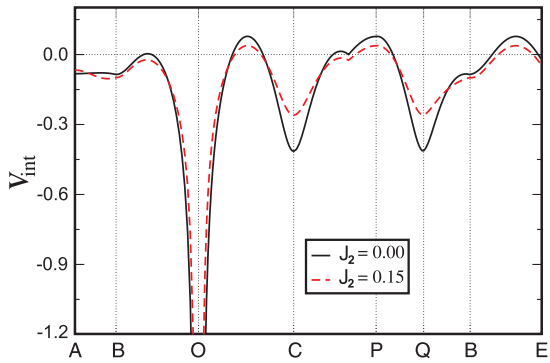}
\caption{Spinon spinon interaction $ V_{\rm int}$ predicted by the SBMF within the first order perturbation theory (Eq. \ref{interact}) for different $J_2$ values and $J^z/J=0.8$. The values of $J_2$ are in units of $J$.}
\label{fig3}
\end{figure}

\noindent If the $XXZ$ Hamiltonian is splitted as $H\!\!=\!\!H_{\rm MF}+V$, the interaction term is given by $V\!\!=\!\!H-H_{\rm MF}$. Then, 
the effect of $V$ on a two free spinon state $ |2{\rm s}\rangle\!\!=\!\!|{\bf q}\uparrow; 
{\bf p} \downarrow\rangle \!\!=\!\!\alpha^{\dagger}_{{\bf q} \uparrow} \alpha^{\dagger}_{
{\bf p}\downarrow}|{\rm gs}\rangle$ is computed, to first order in perturbation theory, as
the energy of creating two spinons above the ground state as $E_{2{\rm s}}\!\!=\!\!\langle2 {\rm s}|H|2 {\rm s}\rangle\!\!-\!\!\langle{\rm gs}|H|{\rm gs}\rangle$.
Therefore, the interaction between the two spinons results $V_{\rm{int}}\!\!=\!\! \overline{E}_{2{\rm s}}-E^{MF}_{2{\rm s}}$, where
$E^{MF}_{2{\rm s}}\!\!=\!\!\langle2 {\rm s}|H_{MF}|2 {\rm s}\rangle-\langle{\rm gs}|H_{MF}|{\rm gs}\rangle\!\!=\omega_{{\bf q}\uparrow}+\omega_{{\bf p}\downarrow}$.\cite{wick} 
 The spinon interaction thus calculated turns out,

\begin{eqnarray}
 V_{\rm{int}} &=&  \frac{1}{N} \big[ \ \gamma_{\bf {q+p}} (u_{ \bf q} v_{\bf p} + v_{ \bf q} u_{ \bf p})^2 + \nonumber \\ 
&& + \frac{J^z}{J} \gamma_{\bf{q-p}} (v_{ \bf q} v_{\bf p} - u_{ \bf q} u_{ \bf p})^2 + \label{interact} \\ 
&& + \frac{J^z}{J} 3 (J+J_{2}) \ \big], \nonumber
\end{eqnarray}

\noindent where $\gamma_{\bf k-\bf p} = \frac{1}{4} \sum_{\bm \delta} J_{\bm\delta} \ e^{ \imath(\bf k-\bf p)\cdot \bm\delta}$. In Fig. \ref{fig3}  is plotted the spinon spinon interaction  $V_{\rm{int}}$ for a pair of spinons $ |\frac{\bf Q}{2}\uparrow;{\bf k}- \frac{\bf Q}{2} \downarrow\rangle \!\!$ building up the lowest magnon excitation of momentum ${\bf k}$, for $J^z/J=0.8$, $J_2=0$ (solid line), and $J_2=0.2$ (dashed line). It is observed that the attraction between two spinons building up the magnon excitation at ${\bf k}=0$ is very strong while for ${\bf k} =\pm{\bf Q}$ the attraction is still, relatively, important. On the other hand, for momenta outside the neighborhood  of ${\bf k}=0$ and ${\bf k} =\pm{\bf Q}$ the attraction of spinon excitations is much weaker. These results agree with the physical picture of tightly bound and weakly bound spinons building up the lower and higher energy magnon excitations, respectively, although within the context of the first order perturbation theory, it is not completely justified. Interestingly, as $J_2$ is introduced there is, in general, a weakening of the spinon spinon interaction for almost all momenta (dashed line of Fig. \ref{fig3}). These results give us a deeper insight of the mean field spectrum. For instance, the spectral weight concentrated at low energy around points C and Q (see $J_2=0$ case of Fig. \ref{fig2}) can be correlated to the presence of tightly bound pairs of spinons building up the magnon excitation; whereas as soon as $J_2$ is increased the spectral weight transfer from low to high energies, along with the appearance of the extended continuum, can be consistently interpreted as the proliferation of nearly free pairs of spinons above the one magnon excitations. 

\begin{figure}[h]
\vspace*{0.cm}
\includegraphics*[width=0.42\textwidth,angle=0]{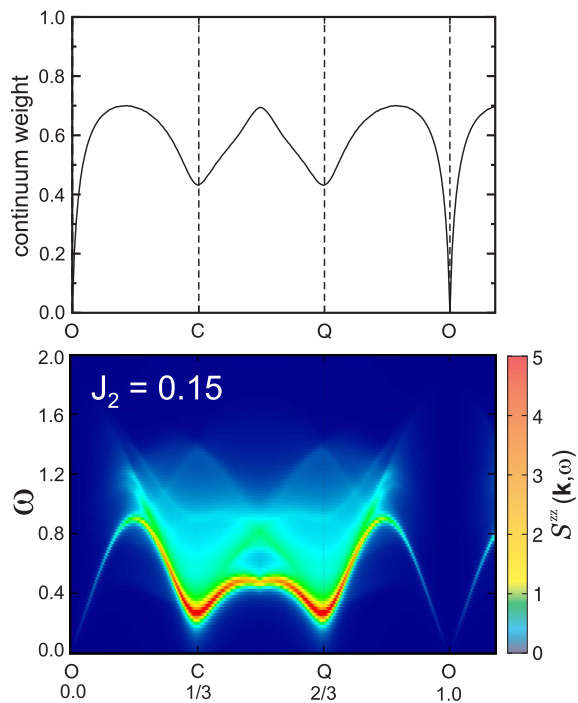}
\caption{Bottom panel: dynamical structure factor predicted by the reconstructed SBMF theory (Intensity curves) along the experimetal path (see Fig. 4(d) of ref.\cite{Zhou12}) for $J^z/J=0.9$. In the horizontal axis it is despicted the cut along [H H 0] direction, in units of $4\pi$, as in ref.\cite{Zhou12} Top panel: relative weight of the two spinon continuum $\int S^{zz}_{cont}({\bf k}, \omega)d\omega/S^{zz}({\bf k}) $. The value of $J_2$ is in units of $J$.}
\label{fig4}
\end{figure}
In order to make a closer comparison with the INS experiments performed in $Ba_3CoSb_2O_9$, in the botton panel of Fig. \ref{fig4} is shown the spectrum predicted by the SBMF theory for $J_2/J=0.15$ and $J^z/J=0.9$ along the experimental path. If one compares with Fig. 4(d) of reference\cite{Zhou12} there is a qualitative good agreement although the dominant high energy spectral weight with respect to the magnon excitation is not completely recovered by the SBMF theory. However, if one separates the spectral weight $S^{zz}_{sing}({\bf k}, \omega)$ of the low energy magnon excitations from the high energy continuum it is possible to quantify the relative weigth of the two spinon continuum in the spectrum by computing 
$\int S^{zz}_{cont}({\bf k}, \omega) d\omega/S^{zz}({\bf k})$ , where $S^{zz}({\bf k})=\int S^{zz}({\bf k},\omega) d\omega$. The top panel of Fig. \ref{fig4} shows an important amount and ${\bf k}$-dependence of the continuum contribution for $J_2=0.15$.\\

\section{Conclusions}

In conclusion, we have performed a series expansion and a mean field Schwinger boson study of the antiferromagnetic $XXZ$ model on the triangular lattice. 
The series expansion results reveal a roton-like excitation minima at the middle points of the edges of the Brillouin zone for all range of anisotropy $0\leq  J^z/J \leq 1$. On the other hand, we have extended the Schwinger boson theory to four bond operators and fully computed static and dynamic properties at the mean field level. The good agreement between the mean field Schwinger boson and the series expansion for the spin wave dispersion relation encouraged us to extend the microscopic model by including exchange interaction to second neighbors in order to qualitatively reproduce the unusual spectrum of the $Ba_3CoSb_2O_9$ compound. By correlating the main features of the mean field spectrum with the spinon spinon interaction we provide a coherent theoretical calculation supporting the idea\cite{Zhou12} that the extended continuum observed in the INS experiments in $Ba_3CoSb_2O_9$ can be interpreted as the fractionalization of magnon excitations in $2D$. Of course, it would be interesting to test the presence of exchange interaction to second neighbors in this compound.  
Another important issue would be to classify the possible spin liquid phases of the $XXZ$ model within a projective symmetry group analysis.\cite{wen02,Messio13,Vishwanath06} Interestingly, using the Schwinger fermions\cite{Reuther14} in the square lattice it has been recently found that the variety of spin liquid phases for a Hamiltonian with $U(1) \times Z_2$ symmetry is even richer than the $SU(2)$ symmetry case.\\

\noindent We thank C. Batista for the exchange of useful information regarding the  $Ba_3CoSb_2O_9$ compound. This work was in part supported by  CONICET (PIP2012) under grant  Nro 1060, by the US National Science Foundation grant number DMR-1306048, and by the computing resources provided by the Australian (APAC) National facility.


\begin{thebibliography}{99}
\bibitem{otros} Using the identities $;\hat{C}^{\dagger}_{ij}\hat{C}_{ij}:+ \hat{D}^{\dagger}_{ij}\hat{D}_{ij}=S^2$ and
$;\hat{B}^{\dagger}_{ij}\hat{B}_{ij}:+ \hat{A}^{\dagger}_{ij}\hat{A}_{ij}=S^2$ for $i\neq j$, the $XXZ$ model can be written in terms of two, three, or four bond operator, being the latter the most apropriate one.
\bibitem{wick} Here the $xy$ part of $E_{2{\rm s}}$ is rescaled as $\overline{E^{xy}}_{2{\rm s}}=\frac{1}{2}E^{xy}_{2{\rm s}}$ in order to compensate the difference, $\langle {\rm gs}|H^{xy}|{\rm gs}\rangle\!\!=\!\!2 \langle{\rm gs}|H^{xy}_{MF}|{\rm gs}\rangle$, between the application of Wick's theorem and mean field decoupling.
\end{thebibliography}
\end{document}